\documentclass[two column, showpacs,preprintnumbers,amsmath,amssymb]{revtex4}
\usepackage{mathrsfs}
\usepackage{graphicx}
\usepackage{dcolumn}
\usepackage{bm}

\begin{document}

\title{The Third Law of Quantum Thermodynamics in the Presence of Anomalous Couplings}

\thanks{Supported by the National Natural Science Foundation
of China under Grant No. 1067401 and the Specialized Research
Foundation for the Doctoral Program of Higher Education under Grant
No. 20050027001.}

\author{WANG Chun-Yang}

\thanks{Corresponding author. Electronic mail: wchy@mail.bnu.edu.cn}

\author{BAO Jing-Dong}

\affiliation{Department of Physics, Beijing Normal University,
Beijing 100875, China}


\begin{abstract}
The quantum thermodynamic functions of a harmonic oscillator
 coupled to a heat bath through velocity-dependent coupling are obtained
 analytically. It is shown that both
the free energy and the entropy decay fast with the temperature in
relation to that of the usual coupling from. This implies that the
 velocity-dependent coupling helps to ensure the third law of
thermodynamics.
\end{abstract}

 \maketitle

The third law of thermodynamics carries prominent consequences for
quantum mechanics and low-temperature physics. It means that all the
thermodynamical quantities vanish when the temperature approaches
the absolute zero. Great progress in the thermodynamics attributed
to this law has been witnessed in elucidating such problems as why
the Carnot engine can never reach $100\%$ efficiency at finite
temperatures. Although some unfulfillment may still exist the known
deviations from the third law will all be cured by quantum
mechanics, quantum statistics and interactions among particles
according to common wisdom. In particular, the recent widespread
interest in the low-temperature behavior of small systems has
highlighted a new angle of viewing the critical role that quantum
dissipative environment plays in a virginal physical field of study,
namely, quantum thermodynamics, for which the validity of the third
law is an unavoidable subject to be elucidated. A rather intriguing
result has been found that a strong  coupling between system and
environment should help to ensure the third law of thermodynamics
\cite{PHan}. This encourages us to consider a further investigation
on the influence of various coupling forms upon the thermodynamical
functions of quantum dissipative system.

Of all  coupling forms of four kinds in the system-plus-reservoir
model, the velocity-dependent coupling, practically exists in
electromagnetic problems such as superconduction quantum
interference device \cite{cl1,cl11,cl2} or blackbody electromagnetic
field \cite{bla}.  This coupling, in the past, is usually believed
to be equivalent to the coordinate-coordinate coupling
\cite{cl1,cl11,cl2,bla,cv1,cv2,cv21}, because the velocity coupling
Hamiltonian can be transformed into a very similar Hamiltonian with
coordinate coupling instead. Nevertheless, the thermal noises
produced by these couplings have different power spectra, especially
at low frequency \cite{hvn2,hvn21,spe}. This provides us a new point
of view to understand the specific character of different
system-reservoir couplings. So it is of great interest to
reinvestigate the third law of thermodynamics of a quantum system
coupled to a heat bath through velocity-dependent coupling.

In this paper, we get the analytical expressions of the free energy
and the entropy of a quantum oscillator in terms of the quantum
generalized Langevin equation, which is easier to be treated than
the quantum propagator approach \cite{bao} for linear problems. It
is shown in all cases that the velocity-dependent coupling exhibits
a great effort to ensure the third law of thermodynamics than the
coordinate-coordinate coupling.

 We start from the generalized Caldeira-Leggett  system-plus-reservoir
  Hamiltonian model \cite{cl1,cl2,cl3} in the
 operator form,
\begin{eqnarray}
\hat{H}&=&\frac{1}{2m}\hat{p}^{2}+\sum^{N}_{j=1}\left[\frac{1}{2m_{j}}(\hat{p}^{2}_{j}+m_{j}^{2}
\omega_{j}^{2}\hat{q}^{2}_{j})+g(\hat{x},\hat{p},\hat{q}_{j},\hat{p}_{j})\right]\nonumber
\\&&+U(\hat{x})
\label{eq:hamiltonian},
\end{eqnarray}
where $\{\hat{x},\hat{p}\}$ and $\{\hat{q}_{j}, \hat{p}_{j}\}$ are
the sets of coordinate and momentum operators of system and bath
oscillators, respectively, for which the commutation relations are
$[\hat{x},\hat{p}]=i\hbar$ and
$[\hat{x}_{i},\hat{p}_{j}]=i\hbar\delta_{ij}$.  The coupling term
$g(\hat{x},\hat{p},\hat{q}_{j},\hat{p}_{j})$ reads
$-c_{j}\hat{x}\hat{q}_{j}+c_{j}^{2}\hat{x}^{2}/(2m_{j}\omega_{j}^{2})$
for the usual coupling between system coordinate and environment
coordinates;
$-d_{1,j}\hat{x}\hat{p}_{j}/m_{j}+d_{1,j}^{2}\hat{x}^{2}/2m_{j}$ or
$-d_{2,j}\hat{p}\hat{q}_{j}/m+d_{2,j}^{2}\hat{q}^{2}_{j}/2m$ for the
system coordinate (momentum) and environment momenta (coordinates)
coupling; and
$-e_{j}\hat{p}\hat{p}_{j}/mm_{j}+e_{j}^{2}\hat{p}_{j}^{2}/(2mm_{j}^{2})$
for momentum-momenta coupling. Noticing that the coupling terms are
so written in order to compensate the coupling induced potential and
mass renormalization.

After eliminating the degrees of freedom of heat bath via the
Heisenberg equations of motion, we get a  quantum generalized
Langevin equation (QGLE)
\begin{eqnarray}
m\ddot{\hat{x}}+\int^{t}_{0}dt'\gamma(t-t')\dot{\hat{x}}(t')+\partial_{\hat{x}}U(\hat{x})=\hat{\xi}(t),\label{eq.GLE}
\end{eqnarray}
where $\gamma(t)$ is the memory friction function and $\hat{\xi}(t)$
is the random force operator with zero mean, its correlation obeys
the quantum fluctuation-dissipation theorem
\cite{nonOhmic1,nonOhmic11}

\begin{eqnarray}
\langle \hat{\xi}(t)\hat{\xi}(t')\rangle_{s}=\frac{\beta\hbar}{\pi
}\int^{\infty}_{0} d\omega
J(\omega)\textrm{coth}(\frac{\beta\hbar\omega}{2})\textrm{cos}(t-t'),
\end{eqnarray}
where $\langle \cdots\rangle_{s}$ denotes the quantum symmetric
average operation and $\beta=1/k_{B}T$ is the inverse temperature.

 In order to  examine the
low-temperature thermodynamical behavior of quantum dissipative
system, we consider a quantum harmonic oscillator:
$U(\hat{q})=\frac{1}{2}m\omega^2_0\hat{q}^2$. By using the
remarkable formula \cite{bla,fre1,fre2,fre21}, we write the free
energy of the quantum oscillator as
\begin{eqnarray}
F(T)=\frac{1}{\pi}\int^{\infty}_{0}d\omega
f(\omega,T)\textrm{Im}\left\{
\frac{d\log\alpha(\omega+i0^{+})}{d\omega}\right\}\label{eq:fre},
\end{eqnarray}
where $f(\omega,T)$ is the free energy of a single oscillator of
frequency $\omega$, given by
$f(\omega,T)=k_{B}T\log[1-\exp(-\hbar\omega/k_{B}T)]$ with the
zero-point contribution $\hbar\omega/2$ being omitted. While
$\alpha(\omega)$ denotes the generalized susceptibility which can be
got from Eq. (\ref{eq.GLE}). Thus the expression of entropy of the
quantum oscillator is given by
\begin{eqnarray}
S(T)=-\frac{\partial F(T)}{\partial T}.\label{eq.entropy}
\end{eqnarray}

Since the function $f(\omega,T)$ in Eq. (\ref{eq:fre}) vanishes
exponentially for $\omega\gg k_{B}T/\hbar$, thus as $T\rightarrow0$
the integrand is confined to low frequencies and we can explicitly
calculate the free energy and then the entropy by expanding the
factor multiplying $f(\omega,T)$ in the powers of $\omega$.

For the harmonic potential, the QGLE is linear and its explicit
solution is obtained
\begin{eqnarray}
\tilde{\hat{x}}(\omega)=\alpha(\omega)\tilde{\hat{\xi}}(\omega),\label{result}
\end{eqnarray}
where
$\tilde{\hat{x}}(\omega)=\int_{-\infty}^{\infty}dt\hat{x}(t)\exp(i\omega
t)$ is the Fourier transform of $\hat{x}(t)$ and similarly noting is
true for $\tilde{\hat{\xi}}(\omega)$. The explicit expression of the
generalized susceptibility in Eq. (\ref{result}) is given by
\begin{eqnarray}
\alpha(\omega)=[-m\omega^{2}-i\omega\tilde{\gamma}(\omega)+m\omega_{0}^{2}]^{-1}\label{eq:alpha}.
\end{eqnarray}

\begin{figure}
\includegraphics[scale=0.6]{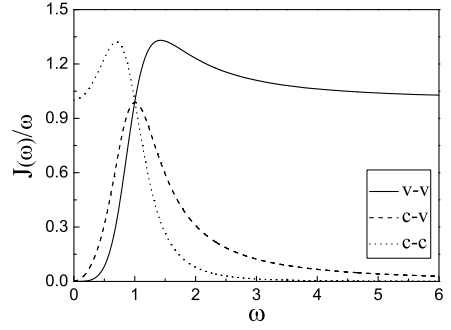}
\caption{\label{fig:fig1}The power spectra of the noise for the
couplings of three kinds: the coordinate-coordinate (c-c),
coordinate-velocity (c-v), and  velocity-velocity (v-v) couplings,
where dimensionless parameters such as $m=1.0$ and
$\gamma_{_{0}}=1.0$ are used as well as $\Gamma=\Omega=1.0$.
\label{spectra}}
\end{figure}

 First, let us consider the usual case of the system's coordinate
 coupled to the coordinates of the heat bath. Experimentally, this type of
 coupling can be realized in a  RLC electric circuit
 driven by a Gaussian white noise.
 Assuming that the power spectrum of the bath oscillators has a
  harmonic form, namely, the spectrum has a narrow Lorentzian distribution
   with the peak centered at a finite frequency. The Fourier
   transform of the memory friction function is
\begin{eqnarray}
\tilde{\gamma}(\omega)=\frac{2m\gamma_{_{0}}\Omega^{4}}
{\Gamma^{2}\omega^{2}+(\Omega^{2}-\omega^{2})^{2}},\label{eq:eq8}
\end{eqnarray}
where $\gamma_{_{0}}$ denotes the  Markovian friction strength of
the system, $\Gamma$ and $\Omega$ are the damping and frequency
parameters of the Gaussian noise, respectively. With this the
response function, Eq. (\ref{eq:alpha}) becomes
\begin{eqnarray}
\alpha(\omega)=\left[-m\omega^{2}-\frac{2i\omega
m\gamma_{_{0}}\Omega^{4}}{\Gamma^{2}\omega^{2}
+(\Omega^{2}-\omega^{2})^{2}}+m\omega_{0}^{2}\right]^{-1},
\end{eqnarray}
and thus the term  appearing in Eq. (4) in the low-frequency limit
is written as
\begin{eqnarray}
&&\textrm{Im}\left\{\frac{d\log\alpha(\omega)}{d\omega}\right\}\nonumber\\
&&\hspace{1cm}=\frac{2\gamma_{_{0}}\Omega^{4}[5\omega^{6}+3(\Gamma^{2}
-(\omega_{_{0}}^{2}+2\Omega^{2})\omega^{4})]}
{(\omega_{_{0}}^{2}-\omega^{2})^{2}(\Gamma^{2}\omega^{2}
+(\Omega^{2}-\omega^{2})^{2})^{2}+4\gamma_{_{0}}^{2}\omega^{2}\Omega^{8}}
\nonumber\\
&&\hspace{1cm}-\frac{2\gamma_{_{0}}\Omega^{4}[(\Gamma^{2}-\Omega^{2}(\Omega^{2}
+2\omega_{_{0}}^{2}))\omega^{2}-\Omega^{4}\omega_{_{0}}^{2}]}
{(\omega_{_{0}}^{2}-\omega^{2})^{2}(\Gamma^{2}\omega^{2}
+(\Omega^{2}-\omega^{2})^{2})^{2}+4\gamma_{_{0}}^{2}\omega^{2}\Omega^{8}}\nonumber\\
&&\hspace{1cm}\cong\frac{2\gamma_{_{0}}}{\omega_{_{0}}^{2}}
\label{eq:hn},
\end{eqnarray}
which reduces to the result of Ohmic friction
\cite{nonOhmic2,non-Ohmic}. Hence, we get the expression of the free
energy of quantum oscillator at low temperature,
\begin{eqnarray}
F(T)&\cong& \frac{2\gamma_{_{0}}k_{B}T}{\pi\omega_{_{0}}^{2}}
\int^{\infty}_{0}d\omega\log[1-\exp(-\hbar\omega/k_{B}T)]\nonumber\\&
=&-\frac{\pi}{3}\hbar\gamma_{_{0}}\left(\frac{k_{B}T}{\hbar\omega_{_{0}}}\right)^{2}.
\end{eqnarray}
The entropy thus reads
\begin{eqnarray}
S(T)=-\frac{\partial F(T)}{\partial
T}=\frac{2\pi}{3}\gamma_{_{0}}\left(\frac{k_{B}^{2}T}{\hbar\omega_{_{0}}^{2}}\right).
\end{eqnarray}
So we have got as $T\rightarrow0$, $S(T)$ vanishes principle to $T$,
in perfect conformity with the third law of thermodynamics and the
linear decay behavior of the entropy is in accordance with the usual
case of Ohmic friction.

Secondly, we consider the anomalous case of the first kind for the
coordinate (velocity) of the system coupled to the velocities
(coordinates) of the heat bath. The spectrum of noise produced by
this coupling
 is shown in Fig. \ref{fig:fig1}. Indeed, it differs very much from
the normal coordinate-coordinate coupling
\cite{hvn1,hvn2,hvn21,hvn11}. The Fourier transform of the
corresponding friction kernel function reads
\begin{eqnarray}
\tilde{\gamma}(\omega)=\frac{2m\gamma_{_{0}}\Gamma^{2}\omega^{2}}
{\Gamma^{2}\omega^{2}+(\Omega^{2}-\omega^{2})^{2}}.\label{eq:eq13}
\end{eqnarray}
In the  limit of low temperature, we have
\begin{eqnarray}
&&\textrm{Im}\left\{\frac{d\log\alpha(\omega)}{d\omega}\right\}\nonumber\\
&&\hspace{1cm}=\frac{4\gamma_{_{0}}\Gamma^{2}\omega[\omega^{4}(\Gamma^{2}
-2(\Omega^{2}-\omega^{2}))+\omega_{_{0}}^{2}(\Omega^{4}-\omega^{4})]}
{(\omega_{_{0}}^{2}-\omega^{2})^{2}(\Gamma^{2}\omega^{2}
+(\Omega^{2}-\omega^{2})^{2})^{2}+4\gamma_{_{0}}^{2}\Gamma^{4}\omega^{4}}\nonumber\\
&&\hspace{1cm}\cong\frac{4\gamma_{_{0}}\Gamma^{2}}{\Omega^{4}\omega_{_{0}}^{2}}\omega
\label{eq:hvn}.
\end{eqnarray}
By using Eq. (14), we get the free energy
\begin{eqnarray}
F(T)&\cong&
\frac{4\gamma_{_{0}}\Gamma^{2}k_{B}T}{\pi\Omega^{4}\omega_{_{0}}^{2}}
\int^{\infty}_{0}d\omega\omega\log[1-\exp(-\hbar\omega/k_{B}T)]\nonumber\\&
=&-\frac{4\gamma_{_{0}}\Gamma^{2}}{\pi\Omega^{4}}
\hbar\omega_{_{0}}\zeta(3)\left(\frac{k_{B}T}{\hbar\omega_{_{0}}}\right)^{3},
\end{eqnarray}
where $\zeta(z)=\Sigma_{n=1}^{\infty}\frac{1}{n^{z}}$ is the
Riemann's zeta-function. This also results in a vanishing entropy in
the low-temperature limit,
\begin{eqnarray}
S(T)=\frac{12\gamma_{_{0}}\Gamma^{2}}{\pi\Omega^{4}}
k_{B}\zeta(3)\left(\frac{k_{B}T}{\hbar\omega_{_{0}}}\right)^{2},
\end{eqnarray}
in agreement with the Nernst's theorem. But the decay behavior of
the entropy as a function of the temperature in the presence of
velocity-dependent coupling is faster than that of the
coordinate-coordinate coupling, as a result of the fact that Eq.
(\ref{eq:hvn}) has a factor $\omega$ whereas the corresponding
result of Eq. (\ref{eq:hn}) is independent of the frequency. This is
due to, from the point of view of theoretical physics, the
velocity-dependent coupling leads the memory friction kernel
function to a strong dependence on the frequency, thus greatly
changed the thermodynamic character of the system.

Finally, let us consider the anomalous dissipation of the second
kind, i.e., the system's velocity is coupled to the volocities of
the heat bath, the spectrum of the noise is also shown in Fig.
\ref{fig:fig1}. The Fourier transform of the friction kernel
function for this coupling reads
\begin{eqnarray}
\tilde{\gamma}(\omega)=\frac{2m\gamma_{_{0}}\omega^{4}}
{\Gamma^{2}\omega^{2}+(\Omega^{2}-\omega^{2})^{2}}.\label{eq:eq17}
\end{eqnarray}
This results in
\begin{eqnarray}
&&\textrm{Im}\left\{\frac{d\log\alpha(\omega)}{d\omega}\right\}\nonumber\\
&&\hspace{0.7cm}=\frac{4\gamma_{_{0}}\omega^{3}[\omega^{6}-\omega^{2}(\Omega^{4}-
(\Gamma^{2}-2\Omega^{2})\omega_{_{0}}^{2})+2\Omega^{4}\omega_{_{0}}^{2}]}
{(\omega_{_{0}}^{2}-\omega^{2})^{2}(\Gamma^{2}\omega^{2}
+(\Omega^{2}-\omega^{2})^{2})^{2}+4\gamma_{_{0}}^{2}\omega^{8}}\nonumber\\
&&\hspace{0.7cm}\cong\frac{8\gamma_{_{0}}}{\Omega^{4}}\omega^{3},
\end{eqnarray}
in the low-frequency, From which we get the expression of the free
energy
\begin{eqnarray}
F(T)=-\frac{48\gamma_{_{0}}}{\pi\Omega^{4}}
\hbar\omega_{_{0}}^{3}\zeta(5)\left(\frac{k_{B}T}{\hbar\omega_{_{0}}}\right)^{5},
\end{eqnarray}
and the entropy is also obtained
\begin{eqnarray}
S(T)=\frac{240\gamma_{_{0}}}{\pi\Omega^{4}}
k_{B}\omega_{_{0}}^{2}\zeta(5)\left(\frac{k_{B}T}{\hbar\omega_{_{0}}}\right)^{4}.
\end{eqnarray}
With this result we conclude again that $S(T)\rightarrow0$ as
$T\rightarrow0$ of no conflicting with the Nernst's theorem, but a
even fast decaying behavior with the temperature for the entropy is
found.

\begin{figure}
\includegraphics[scale=0.6]{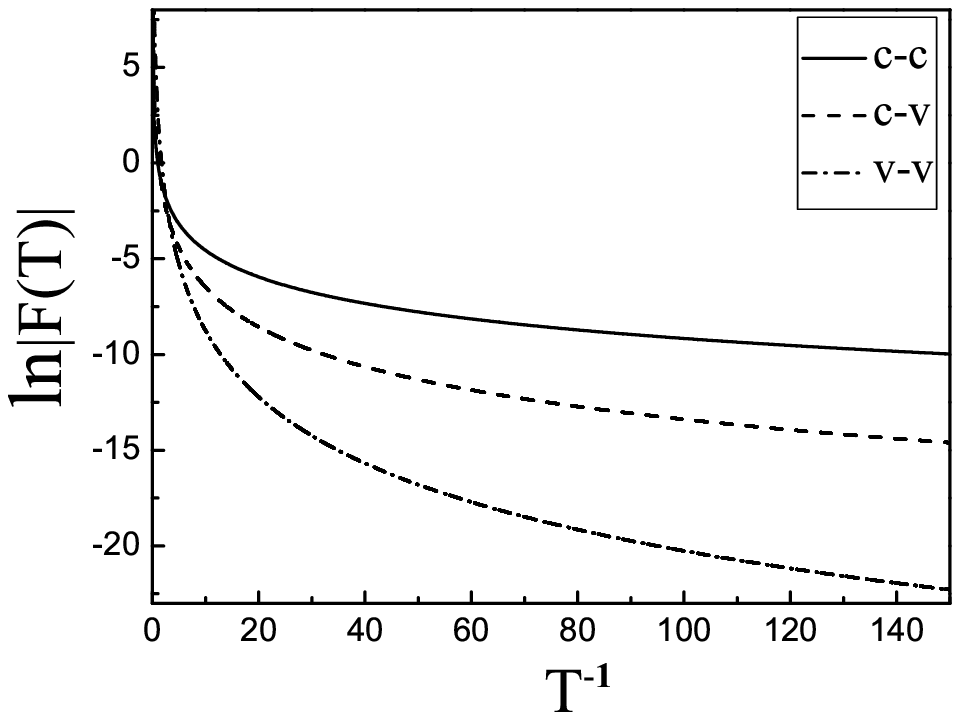}
\includegraphics[scale=0.6]{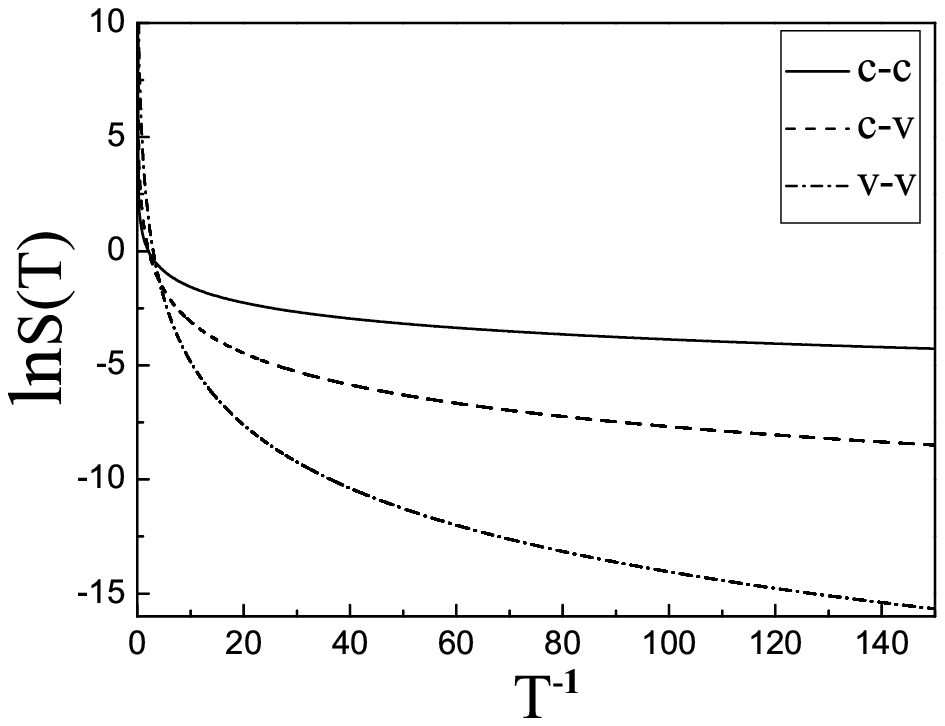}
\caption{\label{fig:fig2}The free energy and the entropy of the
quantum oscillator as functions of the temperature for coupling
forms of three kinds with dimensionless parameters
$\hbar\omega_{_{0}}=k_{B}=1.0$ as well as others used in Fig.
\ref{spectra}.}\label{figure2}
\end{figure}
For the purpose of explicitly distinguishing the decaying behavior
of the free energy and entropies of these three kind of coupling
forms, we give out their log-plot illustration in Fig. \ref{figure2}
as a function of the inverse temperature. From which we can see that
the thermodynamical functions of velocity-dependent coupling systems
exhibit a markedly faster decaying than the coordinate coupling
system. This can be easily understood by comparing Eqs.
(\ref{eq:eq17}) and (\ref{eq:eq13}) with Eq. (\ref{eq:eq8}), where
the friction function of the velocity-dependent coupling is revealed
to be strongly related to the frequency and thus makes the system
more quantum mechanically.

In summary, we have obtained the analytical formula of the free
energy and the entropy of a quantum oscillator coupled to a heat
bath through velocity-dependent coupling form at low temperature.
The low-temperature behavior of the thermodynamical functions has
been discussed. It is shown that the decay behavior of the
thermodynamical functions with the temperature for the anomalous
coupling case is faster than that of the usual coordinate-coordinate
coupling form when the temperature approaches zero. Rather
intriguing is it implied from our study that the fast vanishing
entropies of the velocity-dependent coupling deeply helps to ensure
the validity of the third law of thermodynamics at low temperature.
The results obtained here for the thermodynamical functions of
quantum dissipative system at low-temperature may turn out to be
relevant to experiments in nanoscience where one tests the quantum
thermodynamics of small systems which are coupled to a structured
heat bath. Experimentally, for an example of the system coordinate
(velocity) and environment velocities (coordinates) coupling, one
can study the interaction between a single electron and the
blackbody radiation field where the hamiltonian of the system can be
easily considered under the approximation of dipole polarization.

\end{document}